\RequirePackage[l2tabu, orthodox]{nag}
\documentclass[aps,pra,reprint,superscriptaddress]{revtex4-2}

\usepackage{amsmath,amssymb}
\usepackage{lmodern}

  \usepackage[T1]{fontenc}
  \usepackage[utf8]{inputenc}
  \usepackage{textcomp} 

\usepackage{xcolor}
\usepackage{graphicx}
\usepackage{hyperref}

\begin{document}

\title{Room temperature quantum bit storage exceeding 39 minutes using ionized donors in 28-silicon}

\author{Kamyar Saeedi}
\affiliation{Department of Physics, Simon Fraser University, Burnaby, British Columbia, Canada V5A 1S6}

\author{Stephanie Simmons}
\affiliation{Department of Materials, Oxford University, Oxford, United Kingdom OX1 3PH}

\author{Jeff Z.~Salvail}
\author{Phillip Dluhy}
\affiliation{Department of Physics, Simon Fraser University, Burnaby, British Columbia, Canada V5A 1S6}

\author{Helge Riemann}
\author{Nikolai V. Abrosimov}
\affiliation{Leibniz-Institut f\"{u}r Kristallz\"{u}chtung, 12489 Berlin, Germany}

\author{Peter Becker}
\affiliation{PTB Braunschweig, 38116 Braunschweig, Germany}

\author{Hans-Joachim Pohl}
\affiliation{VITCON Projectconsult GmbH, 07745 Jena, Germany}

\author{John J.~L.~Morton}
\affiliation{London Centre for Nanotechnology, University College
London, United Kingdom WC1H 0AH}

\author{Michael L.~W. Thewalt}
\email[Correspondence to: ]{thewalt@sfu.ca}
\affiliation{Department of Physics, Simon Fraser University, Burnaby, British Columbia, Canada V5A 1S6}

\date{15 Nov 2013}

\begin{abstract}
Quantum memories capable of storing and retrieving
coherent information for extended times at room temperature would enable
a host of new technologies. Electron and nuclear spin qubits using
shallow neutral donors in semiconductors have been studied extensively
but are limited to low temperatures ($\le10$ K); however, the nuclear spins
of ionized donors have potential for high temperature operation. We use
optical methods and dynamical decoupling to realize this potential for
an ensemble of \textsuperscript{31}P donors in isotopically purified
\textsuperscript{28}Si and observe a room temperature coherence time of
over 39 minutes. We further show that a coherent spin superposition can
be cycled from 4.2 K to room temperature and back, and report a
cryogenic coherence time of 3 hours in the same system.
\end{abstract}

\maketitle

A long-term, portable quantum storage register
operating at room temperature would be an important advance in realizing
the potential of quantum computation \cite{r1,r2} and new technologies
such as quantum money \cite{r3,r4}. Solid-state quantum systems have
reached a coherent storage time (\emph{T}\textsubscript{2}) of
\textasciitilde2 s for the nuclear spin of a \textsuperscript{13}C atom
coupled to a nitrogen-vacancy (NV) center in diamond at room temperature
\cite{r5}. Another promising semiconductor qubit system uses the
electron and/or nuclear spins of neutral shallow donor impurities
(D\textsuperscript{0}) such as \textsuperscript{31}P in silicon
\cite{r6,r7,r8}. The nuclear spin of neutral \textsuperscript{31}P in
isotopically purified \textsuperscript{28}Si can reach a coherence time
of 180 s \cite{r9}; however, like all shallow D\textsuperscript{0},
this is an inherently low temperature system. Even at 4.2 K the nuclear
spin \emph{T}\textsubscript{2} is limited by the electron spin
relaxation time, \emph{T}\textsubscript{1} \cite{r9}, which decreases
very rapidly with increasing temperature, dropping to a few milliseconds
at 10 K \cite{r10}; in addition, the donors begin to thermally ionize
above \textasciitilde30 K.

Here we show that the nuclear spin of the ionized donor
(D\textsuperscript{+}) has important advantages over that of
D\textsuperscript{0}, and is not limited to operation at cryogenic
temperatures. In two recent studies on the D\textsuperscript{+} nuclear
spin in natural Si at cryogenic temperatures, one on an ensemble
\cite{r11} and one on a single \textsuperscript{31}P \cite{r12}, the
nuclear spin \emph{T}\textsubscript{2} for D\textsuperscript{+} was
found to be considerably longer than that for D\textsuperscript{0}, as
the removal of the electron spin eliminated decoherence associated with
the electric field noise arising from the nearby electrodes and
Si/SiO\textsubscript{2} interface. The resulting D\textsuperscript{+}
\emph{T}\textsubscript{2} of tens of milliseconds were well accounted
for \cite{r11,r12} by spectral diffusion from the
\textasciitilde5\% of \textsuperscript{29}Si occurring in the natural Si
samples \cite{r13}. Here we remove this source of spectral diffusion by
using highly enriched \textsuperscript{28}Si and dynamic decoupling,

The sample used here and in the previous study of D\textsuperscript{0}
\cite{r9} is enriched to 99.995\% \textsuperscript{28}Si and contains
\textasciitilde5x10\textsuperscript{11} cm\textsuperscript{-3} of
\textsuperscript{31}P and 5x10\textsuperscript{13}
cm\textsuperscript{-3} of the acceptor boron, making it p-type
\cite{r14}. In equilibrium at low temperature one would expect all
donors to be D\textsuperscript{+}, with an equal number of ionized
acceptors, but this equilibrium is reached very slowly at these low
concentrations \cite{r15}. Weak above-gap excitation provided by a 1047
nm laser photoneutralizes almost all of the donors and acceptors. Highly
enriched \textsuperscript{28}Si provides a `semiconductor vacuum' host
for dopants, allowing for the optical hyperpolarization and readout of
D\textsuperscript{0} nuclear spin states \cite{r9,r16}. Here, we
additionally use optical transitions to fully ionize the spin-polarized
D\textsuperscript{0} at low temperature, after which
\emph{T}\textsubscript{1} or \emph{T}\textsubscript{2} measurements can
be carried out on D\textsuperscript{+} either at cryogenic or room
temperature. Following this, with the sample at cryogenic temperature,
the D\textsuperscript{+} are optically reneutralized, and the remaining
D\textsuperscript{0} polarization is read out optically. Once ionized,
virtually all donors will remain ionized indefinitely, independent of
temperature, provided that above-gap light is excluded. Above
\textasciitilde30 K the excess acceptors ionize, providing a background
of free holes, and nearer room temperature thermally generated free
electrons will also be present \cite{r14}. These free carriers could
impact the D\textsuperscript{+} nuclear spin polarization and coherence
times, but our results show that long-term coherent storage at room
temperature is still possible.

\begin{figure}[htbp!]
\includegraphics[width=\columnwidth]{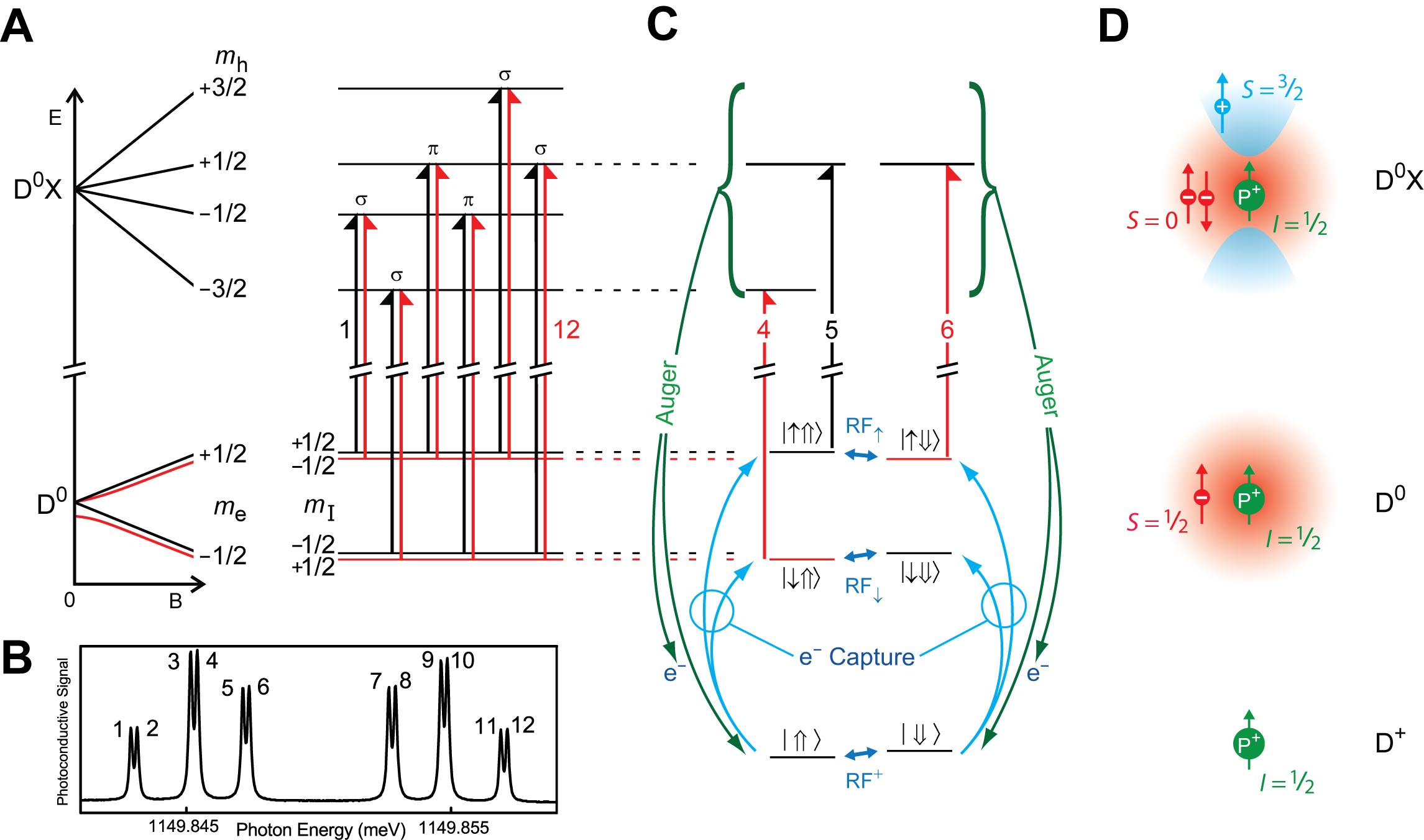}
\caption{\textbf{Energy levels and transitions of the phosphorus neutral donor
(D\textsuperscript{0}), donor bound exciton (D\textsuperscript{0}X) and
ionized donor (D\textsuperscript{+}).} \textbf{(A)} The Zeeman
splittings of the D\textsuperscript{0} and D\textsuperscript{0}X states
are shown from \emph{B}\textsubscript{0}=0 to
\emph{B}\textsubscript{0}=845.3 G, along with the dipole-allowed optical
transitions. \textbf{(B)} Photoconductive readout spectrum without any
D\textsuperscript{0} hyperpolarization. \textbf{(C)} The specific
optical transitions (lines 4, 5, 6) and NMR transitions
(RF\textsubscript{↑}, RF\textsubscript{↓}, RF\textsuperscript{+}) used
here to hyperpolarize, manipulate and read out the nuclear spins. The
magnitude of the D\textsuperscript{+} Zeeman splitting
(RF\textsuperscript{+}) has been exaggerated to show the ordering of the
D\textsuperscript{+} states, and the small nuclear Zeeman energy is
ignored for the D\textsuperscript{0}X states. Note that while the energy
differences between D\textsuperscript{0} and D\textsuperscript{0}X
levels are precisely fixed in \textsuperscript{28}Si, the
D\textsuperscript{+} energy is not well defined due to the kinetic
energy of the e\textsuperscript{-}. \textbf{(D)} Sketches of the spins
and charge densities of D\textsuperscript{+}, D\textsuperscript{0} and
D\textsuperscript{0}X.
\label{fig1}}
\end{figure}

\begin{figure}[htbp!]
\includegraphics[width=\columnwidth]{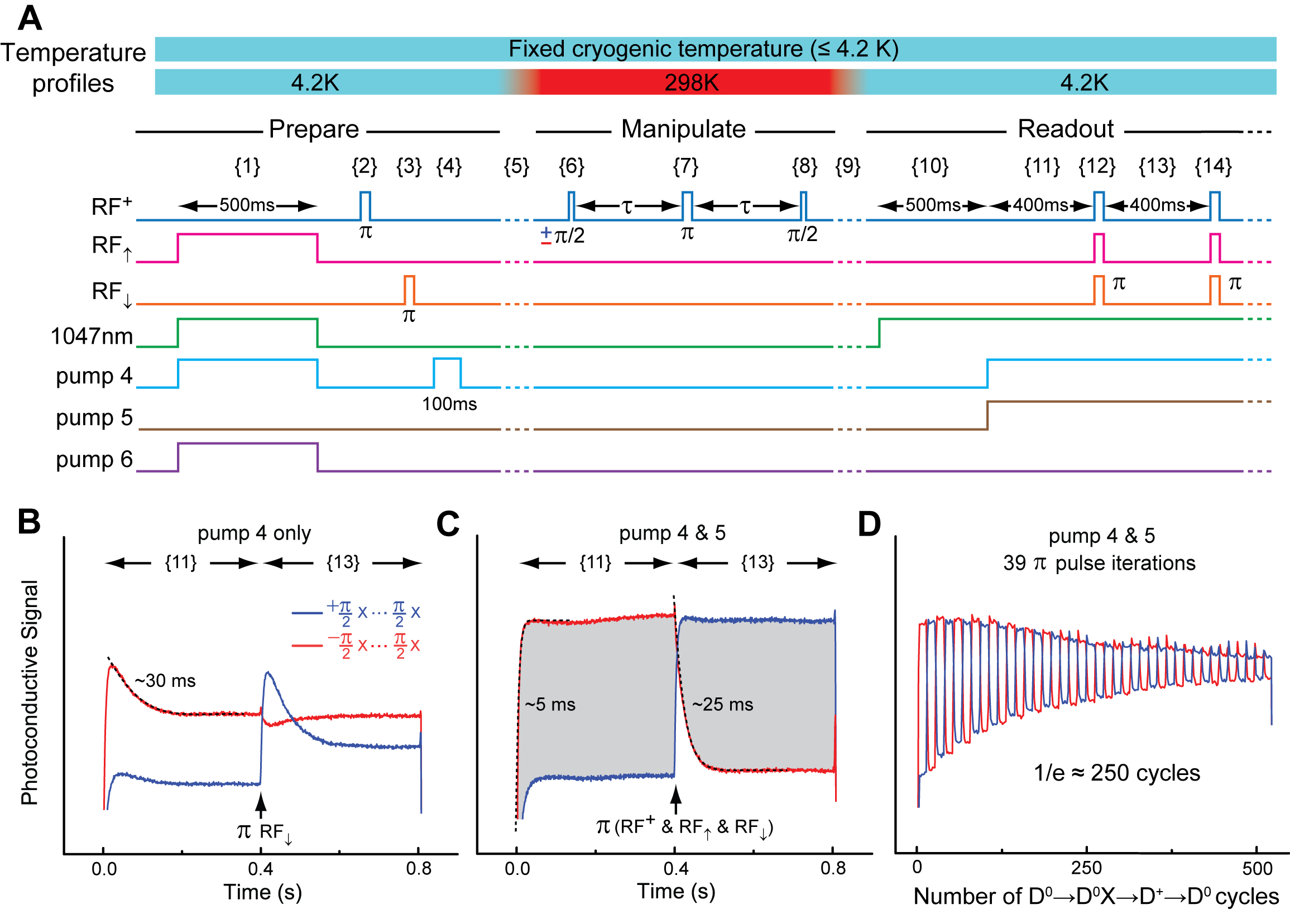}
\caption{\textbf{Initialization, manipulation and readout protocols.}
\textbf{(A)} The laser and RF sequences used to prepare
D\textsuperscript{+} in the $\lvert\Uparrow\rangle$ state (\{1\} -- \{4\}),
manipulation of D\textsuperscript{+} spins for the case of a Hahn echo
(\{6\} -- \{8\}), and readout of the resulting Z-component (\{10\} --
\{14\}\ldots). At top are the two temperature profiles relevant to Fig.~\ref{fig3}: either a constant temperature $\le$ 4.2~K, or 4.2~K during preparation
and readout, with a ramp up to 298~K taking \textasciitilde{} 6 minutes
\{5\}, a constant 298~K during the D\textsuperscript{+} manipulation
period, and a ramp down to 4.2~K taking \textasciitilde4 minutes \{9\}.
Each measurement is performed twice, with opposite signs of the initial
$\pi{}/2$ pulse \{6\}. \textbf{(B)} Single-shot readout of
D\textsuperscript{+} polarized $\lvert\Uparrow\rangle$ (red) or $\lvert\Downarrow\rangle$ (blue)
using our previous method optimized for D\textsuperscript{0} readout is
compared with \textbf{(C)}, the improved scheme for D\textsuperscript{+}
readout \cite{r14}. The detected signal is proportional to the shaded
area. \textbf{(D)} The cycle shown in \textbf{(C)} extended to $39\pi$
pulse inversions (16 s).
\label{fig2}}
\end{figure}

The optical transitions between D\textsuperscript{0} and the donor bound
exciton (D\textsuperscript{0}X) used for hyperpolarization, readout and
donor ionization are shown in Fig.~\ref{fig1}. In Fig.~\ref{fig1}C, the four
D\textsuperscript{0} hyperfine levels are labeled by their electron spin
($\uparrow$ or 	$\downarrow$) and nuclear spin ($\Uparrow$ or $\Downarrow$) (the $\lvert\uparrow\Downarrow\rangle$ and  $\lvert\downarrow\Uparrow\rangle$
labels are approximate at low \emph{B}\textsubscript{0} because of
hyperfine mixing). The D\textsuperscript{0}X decay with near-unity
efficiency through the Auger process \cite{r17} to give
D\textsuperscript{+} and free electrons (e\textsuperscript{-}), which
are eventually recaptured to return D\textsuperscript{+} to
D\textsuperscript{0}. The Auger decay process is central to both the
D\textsuperscript{0} hyperpolarization and hyperfine state readout using
resonant D\textsuperscript{0}X photoconductivity, as illustrated in the
sequence used to measure the coherence time of D\textsuperscript{+}
nuclear spins (see Fig.~\ref{fig2}A, and \cite{r14}. It consists of optical and
RF pulses to hyperpolarize the nuclear spins (Steps \{1-3\}), fully
ionize the donors \{4\}, coherently manipulate the D\textsuperscript{+}
nuclear spin \{6-8\}, reneutralize the donors \{10\}, and read out the
resulting spin populations \{11-14\}. By step \{5\} we estimate that
over 90\% of the \textsuperscript{31}P are both ionized and polarized
into $\lvert\Uparrow\rangle$. Single shot readouts of D\textsuperscript{+},
polarized into either $\lvert\Uparrow\rangle$ or $\lvert\Downarrow\rangle$ and then reneutralized,
are shown in Fig.~\ref{fig2}B-D, contrasting our previous \cite{r9} readout
method optimized for D\textsuperscript{0} (Fig.~\ref{fig2}B) with the improved
readout used here (Fig.~\ref{fig2}C-D). Details of the preparation and readout
schemes are found in \cite{r14}. Analysis of the data in Fig.~\ref{fig2}C and D
shows that the D\textsuperscript{0} → D\textsuperscript{0}X →
D\textsuperscript{+} → D\textsuperscript{0} readout cycle can be
repeated at least 250 times before the nuclear polarization decays by
1/e \cite{r14}, which is an underestimate given that much of the decay
in Fig.~\ref{fig2}D is due to imperfections in the readout $\pi$ pulses. A similar
insensitivity of the nuclear spin polarization to repeated donor charge
cycles has been reported for readout of a single \textsuperscript{31}P
nuclear spin \cite{r12}, and for ensemble measurements using
electrically detected magnetic resonance \cite{r20}.

We use two different temperature profiles to measure
\emph{T}\textsubscript{1} and \emph{T}\textsubscript{2}, as shown above
Fig.~\ref{fig2}A, either a fixed temperature at or below 4.2 K, or
\emph{T}\textsubscript{1} or \emph{T}\textsubscript{2} measurements at
room temperature (298 K) with the polarization and readout steps at 4.2
K. The measurement RF pulse sequence is shown for a simple Hahn echo
($\pi{}/2$ - $\pi$ - $\pi{}/2$). The temperature is changed only while the
D\textsuperscript{+} nuclear spin is in an eigenstate in the Z-basis
(i.e. in the $\lvert\Downarrow\rangle$ or $\lvert\Uparrow\rangle$ state). This ensures the nuclear
spin is sensitive only to \emph{T}\textsubscript{1} relaxation processes
while the temperature is changing. Later we explore a third profile,
changing the temperature while the nuclear spin is in a superposition
state.

\begin{figure}[htbp!]
\includegraphics[width=\columnwidth]{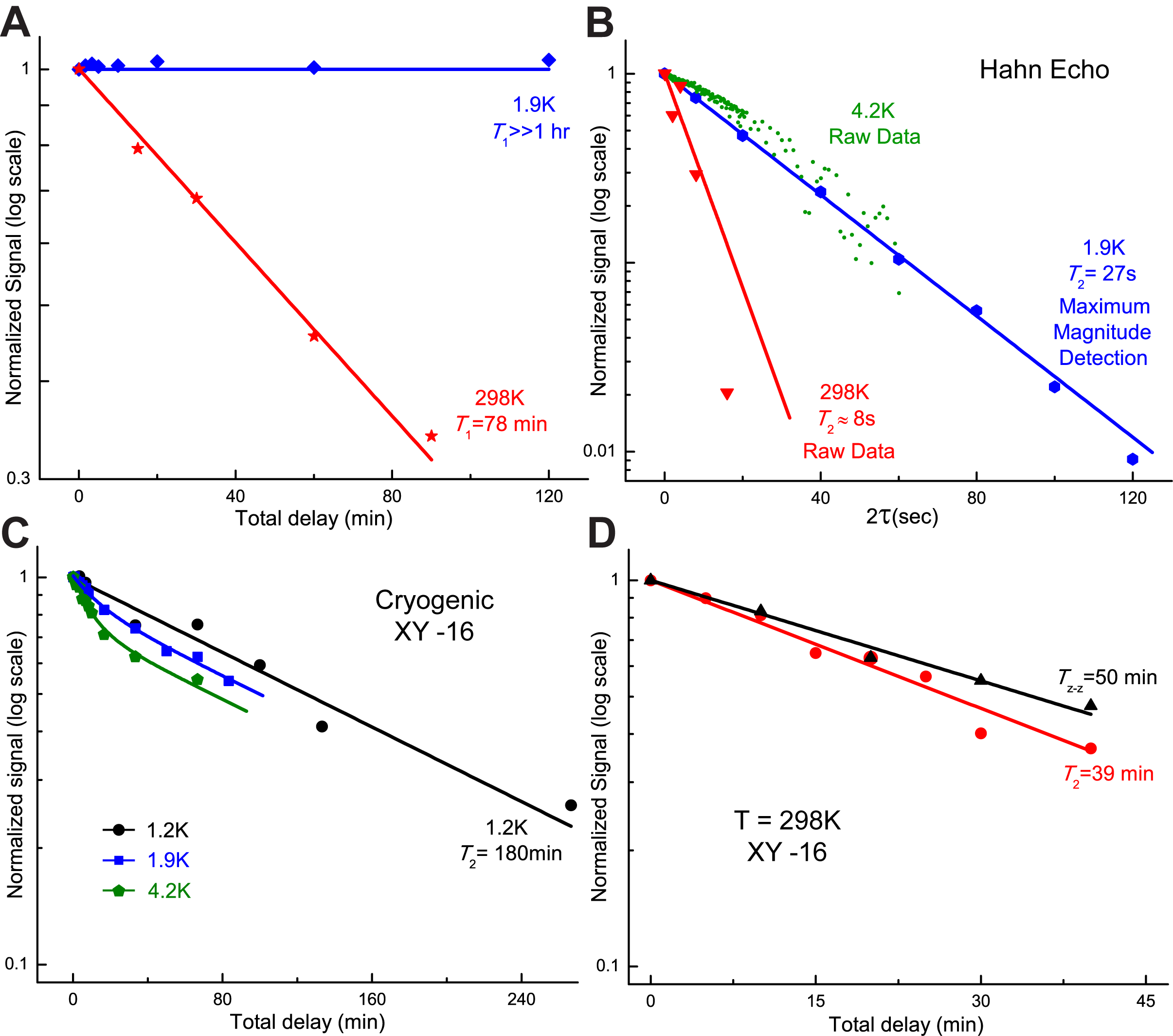}
\caption{\textbf{Measured \emph{T}\textsubscript{1} and \emph{T}\textsubscript{2}
times for the \textsuperscript{31}P\textsuperscript{+} nuclear spin at
cryogenic and room temperature.} \textbf{(A)} The decay of the nuclear
spin polarization (along Z), parameterized by \emph{T}\textsubscript{1},
is shown for 1.9~K and 298~K. \textbf{(B)~}Single shot Hahn echo
\emph{T}\textsubscript{2} measurements are shown at 298~K and 4.2~K, the
latter (green dots) showing increasing phase noise with increasing
delay. The effect of phase noise can be suppressed by using maximum
magnitude detection (\emph{14}), as shown for data taken at 1.9~K.
\textbf{(C)} \emph{T}\textsubscript{2} decays using the XY-16 decoupling
sequence at cryogenic temperatures. The 1.9 K and 4.2 K data are fit
using biexponentials, with the longer component set to 180 min.
\textbf{(D)} The \emph{T}\textsubscript{2} decay at 298~K using XY-16
decoupling, together with the observed decay of a ±Z state using XY-16
decoupling under identical conditions.
\label{fig3}}
\end{figure}

In Fig.~\ref{fig3}A we show the D\textsuperscript{+} nuclear spin
\emph{T}\textsubscript{1} measured at 1.9 K and room temperature (the
Hahn echo sequence is replaced with either no operations, leaving the
nuclear spin polarization unchanged, or a $\pi$ pulse, which inverts it
\cite{r14}). The D\textsuperscript{+} \emph{T}\textsubscript{1} at
cryogenic temperature was so long that no decay could be observed over 2
hours, and at room temperature \emph{T}\textsubscript{1} is over an
hour. We note that even a short thermal cycle up to room temperature and
back resulted in a \textasciitilde30\% loss in nuclear spin polarization
compared to the same measurement at a constant 4.2 K, so all room
temperature decay data is normalized to unity for the shortest time (2
min. at 298 K). Fig.~\ref{fig3}B shows single-shot Hahn echo decay data at 4.2 K
revealing increasing phase noise with increasing delay time, likely
arising from low-frequency magnetic field fluctuations. This phase noise
was eliminated from the 1.9 K data by using maximum magnitude detection
\cite{r14}. The Hahn echo \emph{T}\textsubscript{2} of about 30 s
measured at or below 4.2~K is well explained by spectral diffusion due
to the residual (46 ppm) \textsuperscript{29}Si nuclear spins present in
the sample \cite{r13}. Also shown is single-shot Hahn echo data at room
temperature, where the long cycle time made the use of maximum magnitude
detection impractical, so that the phase noise could not be eliminated
and the apparent Hahn echo \emph{T}\textsubscript{2} is reduced to
\textasciitilde8 s.

We have demonstrated \cite{r9} that dynamic decoupling using the XY-16
sequence of $\pi$ pulses \cite{r19} is effective for reducing the effect of
low frequency noise on donor nuclear spins while maintaining arbitrary
initial states. In Fig.~\ref{fig3}C we show the results of using this sequence to
replace the single $\pi$ pulse of the Hahn echo (for all XY-16 results shown
here the time $2\tau$ between $\pi$ pulses was 8 ms). At 1.2 K the coherence
decay follows a single exponential with a \emph{T}\textsubscript{2} of
180 min., whereas at 1.9 K and 4.2 K there is an early component of a
faster decay (time constant \textasciitilde{} 12 min.) followed by a
decay consistent with a \emph{T}\textsubscript{2} of 180 min. We believe
that this initial faster decay is due to charge dynamics in the sample
after illumination, likely from D\textsuperscript{-} and
A\textsuperscript{+} centers, which are frozen out at the lowest
temperature \cite{r20}. It may be related to the \textasciitilde30\%
loss in nuclear polarization observed in even short cycles from
cryogenic to room temperature and back. In Fig.~\ref{fig3}D we show a room
temperature \emph{T}\textsubscript{2} decay of 39 min. This is a lower
bound, because the same XY-16 sequence applied to a ±Z state yields a
decay constant of 50 min., substantially shorter than the 78 min.
\emph{T}\textsubscript{1}, indicating that pulse errors in the XY-16
sequence contribute significantly to the observed decay, and are also
likely to contribute to the 180 min. \emph{T}\textsubscript{2} observed
at cryogenic temperatures.

The low-temperature nuclear spin \emph{T}\textsubscript{2} of $\ge180$~min.
demonstrates that the XY-16 sequence is very effective in suppressing
decoherence arising from slow spectral diffusion caused by the remaining
\textsuperscript{29}Si. It is interesting to note that whereas the
cryogenic Hahn echo \emph{T}\textsubscript{2} reported here for
D\textsuperscript{+} is slightly shorter than that reported earlier
\cite{r10} for D\textsuperscript{0}, XY-16 dynamic decoupling extends
the observed coherence time by a factor of 400 for D\textsuperscript{+},
but only by \textasciitilde4.4 for D\textsuperscript{0}. This suggests a
very different decoherence process for the D\textsuperscript{0} case,
\cite{r14}.

These long coherence times for the D\textsuperscript{+} nuclear spin
should be achievable even when the donor is placed near an interface in
a nanodevice, as long as the temperature is low enough that flips or
flip-flops of electron spins at the interface are suppressed. The
shorter 39 min. \emph{T}\textsubscript{2} measured at room temperature
could arise from carrier-induced magnetic field fluctuations, whose
effect is not completely suppressed by the dynamical decoupling,
combined with a higher error in the RF pulses \cite{r15}. The observed
room temperature \emph{T}\textsubscript{2} is also compatible with the
accumulated phase error from the small probability of the donor being in
the D\textsuperscript{0} ground state at room temperature. The observed
room temperature \emph{T}\textsubscript{2} considerably exceeds that
reported \cite{r21} for \textsuperscript{29}Si in natural Si using
homonuclear decoupling. Given that \textsuperscript{29}Si should not be
more sensitive to free carriers than D\textsuperscript{+}, this likely
results from difficulty in completely decoupling the
\textsuperscript{29}Si at the high concentration present in natural Si.

\begin{figure}[htbp!]
\includegraphics[width=\columnwidth]{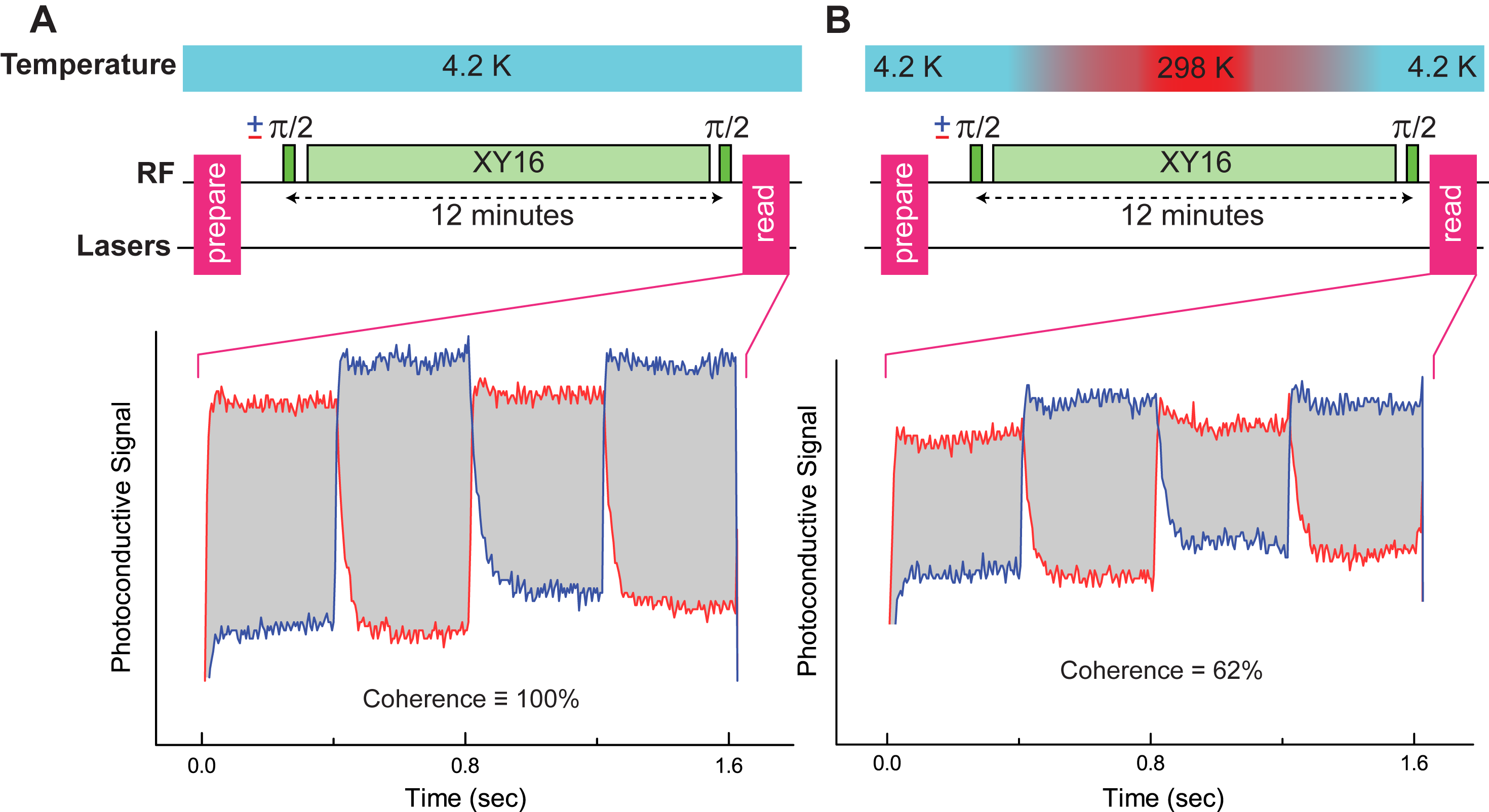}
\caption{\textbf{Cycling D\textsuperscript{+}, while in a nuclear spin
superposition state, from 4.2~K to room temperature and back.}
\textbf{(A)} A measurement at a constant temperature of 4.2~K, with
XY-16 decoupling over a 12 min. period, is compared to \textbf{(B)}
where the nuclear spins are placed into a coherent superposition at
4.2~K and the XY-16 decoupling sequence is begun, followed by a
\textasciitilde6 min. ramp to 298~K, 2 min. at 298~K, and a
\textasciitilde4 min. ramp back down to 4.2~K, after which the remaining
coherence is read out. The preparation and readout sequences are as in
Fig. 2A. A comparison of \textbf{(A)} and \textbf{(B)} shows that 62\%
of the spin coherence remains after the temperature cycle, equivalent to
a state fidelity of 81\%.
\label{fig4}}
\end{figure}

Finally, we demonstrate the ability to change the sample temperature
while the D\textsuperscript{+} nuclear spin is in a coherent
superposition state. Fig.~\ref{fig4}A shows a reference measurement at 4.2 K
using the sequence shown in Fig.~\ref{fig2}A, but with XY-16 decoupling. In Fig.~\ref{fig4}, the D\textsuperscript{+} nuclear spins are placed into a coherent
superposition at 4.2 K, the XY-16 sequence is begun, and then the
temperature is ramped to room temperature in \textasciitilde6 min. It is
held there for 2 min. before being ramped back down to 4.2~K in
\textasciitilde4 min. Once the sample is reimmersed in liquid He, the
XY-16 sequence ends and the remaining coherence is projected back into a
±Z state for readout after reneutralization. By comparing the two
readout signals we see that it is possible to bring a coherent state
from cryogenic temperature to room temperature and back while retaining
62\% of the coherence signal, which is equivalent to a state fidelity of
81\% \cite{r22}. This loss of coherence can be largely attributed to
the \textasciitilde30\% drop in nuclear spin polarization observed over
one thermal cycle to room temperature and back.

These results support the possibility of truly long term storage of
quantum information at room temperature. To make use of the
D\textsuperscript{+} state as a quantum memory for, say, a donor-based
electron spin qubit, as has already been done with the nuclear spin of
D\textsuperscript{0} \cite{r23}, it will be necessary to find a way to
ionize and neutralize the donor without disturbing the coherent state of
the nuclear spin. Whereas \textsuperscript{31}P donors in
\textsuperscript{28}Si at this time require low temperatures for
initialization and readout, the ability to bring coherent information
reversibly between cryogenic and room temperatures already suggests ways
to exploit this system. It may also be possible to initialize and read
out this system at elevated temperatures, or to find similar but more
robust systems with larger electron binding energies, in which charge
control can still be used to turn a hyperfine interaction on for
initialization and readout and off for long term storage. In Si, one
possibility would be to use much deeper donors such as chalcogens, where
an optically accessible hyperfine splitting has already been observed
for \textsuperscript{77}Se\textsuperscript{+} in \textsuperscript{28}Si
\cite{r24}, and where the hyperfine coupling can be removed by placing
the donor into either D\textsuperscript{0} or D\textsuperscript{2+}
charge states. Another promising possibility would be deep defects in
wider-gap materials such as diamond and SiC \cite{r25}, which can also
be isotopically purified to remove background spins, and where the
method of charge state control could be combined with initialization and
readout at room temperature.

\section*{Acknowledgments} The work at SFU was supported by the Natural
Sciences and Engineering Research Council of Canada (NSERC). S.S. is
supported by the Violette and Samuel Glasstone Fellowship and St.
John\textquotesingle s College, Oxford. J.J.L.M. is supported by the
Royal Society.

\section*{Supplementary Materials}

Materials and Methods

Supplementary Text

Figures S1-S3

References \cite{r26,r27,r28,r29,r30,r31,r32,r33,r34,r35,r36,r37}


\begin{thebibliography}{99}

\bibitem{r1} D. Deutsch, Quantum theory, the Church-Turing principle and the
universal quantum computer. \emph{Proc. R. Soc. Lond. A} \textbf{400}, 97--117 (1985).

\bibitem{r2} T. D. Ladd \emph{et al}., Quantum computers. \emph{Nature}
\textbf{464}, 45-53 (2010).

\bibitem{r3} S. Wiesner, Conjugate coding. \emph{ACM SIGACT News} \textbf{15}, 78-88 (1983).

\bibitem{r4} Fernando Pastawski, Norman Y.~Yao, Liang Jiang, Mikhail D.~Lukin, J.~Ignacio Cirak, Unforgeable noise-tolerant quantum tokens. \emph{PNAS} \textbf{109}, 16079-16082 (2012).

\bibitem{r5} P. C. Maurer \emph{et al}., Room-temperature quantum bit memory
exceeding one second. \emph{Science} \textbf{336}, 1283-1286 (2012).

\bibitem{r6} B. E. Kane, A silicon-based nuclear spin quantum computer.
\emph{Nature} \textbf{393}, 133-137 (1998).

\bibitem{r7} John J. L. Morton, Dane R. McCamey, Mark A. Eriksson, Stephen A.
Lyon, Embracing the quantum limit in silicon computing. \emph{Nature} \textbf{479}, 345-353
(2011).

\bibitem{r8} David D. Awschalom, Lee C. Bassett, Andrew S. Dzurak, Evelyn L. Hu,
Jason R. Petta,
Quantum spintronics: engineering and manipulating atom-like spins in semiconductors.
\emph{Science} \textbf{339}, 1174-1179 (2013).

\bibitem{r9} M. Steger \emph{et al}., Quantum information storage for over 180 s
using donor spins in a \textsuperscript{28}Si semiconductor vacuum. \emph{Science} \textbf{336}, 1280-1283 (2012).

\bibitem{r10} G. Feher and E. A. Gere, Electron spin resonance experiments on
donors in silicon. II. electron spin relaxation effects. \emph{Phys. Rev.} \textbf{114},
1245-1256 (1959).

\bibitem{r11} Lukas Dreher, Felix Hoehne, Martin Stutzmann, Martin S. Brandt,
Nuclear spins of ionized phosphorus donors in silicon. \emph{Phys. Rev. Lett}. \textbf{108},
027602 (2012).

\bibitem{r12} Jarryd J. Pla \emph{et al}., High-fidelity readout and control of a
nuclear spin qubit in silicon. \emph{Nature} \textbf{496}, 334-338 (2013).

\bibitem{r13} Wayne M. Witzel, Malcolm S. Carroll, Łukasz Cywiński, S. Das Sarma,
Quantum decoherence of the central spin in a sparse system of dipolar coupled
spins. \emph{Phys. Rev. B} \textbf{86}, 035452 (2012).

\bibitem{r14} Supplementary materials are available on \emph{Science} Online.

\bibitem{r15} P. Dirksen, A. Henstra, W. Th. Wenckebach, An electron spin-echo
study of donor-acceptor recombination. \emph{J. Phys. Condens. Matter} \textbf{1}, 7085 (1989).

\bibitem{r16} M. Steger \emph{et al.,} Optically-detected NMR of
optically-hyperpolarized \textsuperscript{31}P neutral donors in \textsuperscript{28}Si\emph{. J. Appl. Phys.} \textbf{109}, 102411
(2011).

\bibitem{r17} W. Schmid, Auger lifetimes for excitons bound to neutral donors and
acceptors in Si\emph{. Phys. Stat. Sol. (b)} \textbf{84}, 529-540 (1977).

\bibitem{r18} D. R. McCamey, J. Van Tol, G. W. Morley, C. Boehme, Electronic spin
storage in an electrically readable nuclear spin memory with a lifetime
\textgreater100 seconds. \emph{Science} \textbf{330}, 1652-1656 (2010).

\bibitem{r19} T. Gullion, D. B. Baker and M. S. Conradi, New, compensated
Carr-Purcell sequences. \emph{J. Magn. Reson.} \textbf{89}, 479-484 (1990).

\bibitem{r20} W. Burger and K. Lassmann, Energy-resolved measurements of the
phonon-ionization of \emph{D}\textsuperscript{-} and \emph{A}\textsuperscript{+} centers in silicon with
superconducting-Al tunnel junctions. \emph{Phys. Rev. Lett}.
\textbf{53}, 2035-2037 (1984).

\bibitem{r21} T. D. Ladd, D. Maryenko, Y. Yamamoto, E. Abe, K. M. Itoh, Coherence
time of decoupled nuclear spins in silicon. \emph{Phys. Rev. B} \textbf{71}, 014401
(2005).

\bibitem{r22} R. Jozsa, Fidelity for mixed quantum states. \emph{J. Modern Optics}
\textbf{41}, 2315-232 (1994).

\bibitem{r23} J. J. L. Morton \emph{et al}., Solid-state quantum memory using the
\textsuperscript{31}P nuclear spin\emph{. Nature} \textbf{455}, 1085-1088 (2008).

\bibitem{r24} M. Steger \emph{et al}., High-resolution absorption spectroscopy of
the deep impurities S and Se in \textsuperscript{28}Si revealing the \textsuperscript{77}Se hyperfine
splitting. \emph{Phys. Rev. B} \textbf{80}, 115204, (2009).

\bibitem{r25} W. F. Koehl, B. B. Buckley, F. J. Heremans, G. Calusine, D. D.
Awschalom, Room temperature coherent control of defect spin qubits in silicon carbide.
\emph{Nature} \textbf{479}, 84 (2011).

\bibitem{r26} P. Becker, H.-J. Pohl, H. Riemann and N. V. Abrosimov, Enrichment of
silicon for a better kilogram. \emph{Phys. Stat. Sol. (a)} \textbf{207}, 49-66 (2010).

\bibitem{r27} T. Sekiguchi \emph{et al}., Hyperfine structure and nuclear
hyperpolarization observed in the bound exciton luminescence of Bi donors in natural Si. \emph{Phys. Rev. Lett}.
\textbf{104}, 137402 (2010).

\bibitem{r28} L. Viola and S. Lloyd, Dynamical suppression of decoherence in
two-state quantum systems. \emph{Phys. Rev. A} \textbf{58}, 2733-44 (1998).

\bibitem{r29} A. M. Tyryshkin \emph{et al}., Effect of pulse error accumulation on
dynamical decoupling of the electron spins of phosphorus donors in silicon. \emph{Phys. Rev. B}
\textbf{85}, 085206 (2012).

\bibitem{r30} S. Wimperis, Broadband, narrowband, and passband composite pulses
for use in advanced NMR experiments. \emph{J. Magn. Reson. Ser. A} \textbf{109}, 221-231
(1994).

\bibitem{r31} A. M. Tyryshkin \emph{et al}., Electron spin coherence exceeding
seconds in high-purity silicon. \emph{Nature Mat.} \textbf{11}, 143-147 (2012).

\bibitem{r32} G. Wolfowicz \emph{et al}., Atomic clock transitions in
silicon-based spin qubits, \emph{Nature Nanotech}. \textbf{8}, 561-564 (2013).

\bibitem{r33} F. R. Bradbury \emph{et al}., Stark tuning of donor electron spins
in silicon, \emph{Phys. Rev. Lett}. \textbf{97}, 176404 (2006).

\bibitem{r34} M. Friesen, Theory of the Stark effect for P donors in Si,
\emph{Phys. Rev. Lett}. \textbf{94}, 186403 (2005).

\bibitem{r35} L. Dreher \emph{et al}., Electroelastic hyperfine tuning of
phosphorus donors in silicon, \emph{Phys. Rev. Lett}. \textbf{106}, 037601 (2011).

\bibitem{r36} D. Karaiskaj \emph{et al}., Origin of the residual acceptor ground
state splitting in silicon, \emph{Phys. Rev. Lett}. \textbf{90}, 016404 (2003).

\bibitem{r37} D. M. Larsen, Inhomogeneous broadening of the Lyman-series
absorption of simple hydrogenic donors, \emph{Phys. Rev. B} \textbf{13}, 1681-1691 (1978).

\end{thebibliography}
\end{document}